\documentstyle[11pt]{article}
\def\bsh{\backslash}
\def\bdt{\dot \beta}
\def\adt{\dot \alpha}

\newfont{\bbbold}{msbm10 scaled \magstep1}

\def\bbM{\mbox{\bbbold M}}

\def\bbZ{\mbox{\bbbold Z}}
\def\cA{\cal A}

\def\cS{\cal S}
\def\cT{\cal T}

\def\cV{{\cal V}}

\newfont{\goth}{eufm10 scaled \magstep1}

\def\gl{\mbox{\goth l}}

\def\gp{\mbox{\goth p}}

\def\gs{\mbox{\goth s}}

\def\gu{\mbox{\goth u}}

\def\a{\alpha}
\def\b{\beta}

\def\d{\delta}\def\D{\Delta}
\def\e{\epsilon}

\def\l{\lambda}\def\L{\Lambda}
\def\m{\mu}

\def\p{\pi}

\def\t{\tau}
\def\th{\theta}

\def\be{\begin{equation}}\def\ee{\end{equation}}
\def\bea{\begin{eqnarray}}\def\eea{\end{eqnarray}}
\def\ba{\begin{array}}\def\ea{\end{array}}

\def\del{\partial}

\def\sd{\rm sdet}\def\str{\rm str}
\def\xz{\times}

\def\nab{\nabla}

\def\del{\partial}

\let\la=\label

\let\bm=\bibitem

\def\bd{\begin{document}}
\def\ed{\end{document}}
\def\bea{\begin{eqnarray}}
\def\eea{\end{eqnarray}}
\def\ba{\begin{array}}
\def\ea{\end{array}}
\def\ft#1#2{{\textstyle{{\scriptstyle #1}\over {\scriptstyle #2}}}}
\def\fft#1#2{{#1 \over #2}}
\newcommand{\eq}[1]{(\ref{#1})}
\def\eqs#1#2{(\ref{#1}-\ref{#2})}
\def\det{{\rm det\,}}
\def\tr{{\rm tr}}
\newcommand{\ho}[1]{$\, ^{#1}$}
\newcommand{\hoch}[1]{$\, ^{#1}$}
\newcommand{\tamphys}{\it\small Center for Theoretical Physics,
Texas A\&M University, College Station, TX 77843, USA}
\newcommand{\newton}{\it\small Isaac Newton Institute for Mathematical
Sciences, Cambridge, UK}
\newcommand{\kings}{\it\small Department of Mathematics, King's College,
London, UK}

\begin{document}


\hfill{hep-th/9905085}

\hfill{\today}

\vspace{30pt}

\begin{center}
{\Large{\bf Nilpotent invariants in $N=4$ SYM}}
\vspace{30pt}

B. Eden, P.S. Howe and P.C. West
\vskip 30pt
Department of Mathematics
King's College
London UK

\vspace{30pt}

{\bf Abstract}

\end{center}

It is shown that there are no
nilpotent invariants in $N=4$ analytic superspace for $n\leq4$ points.
It is argued that there is (at least) one such invariant for
$n=5$ points which is not invariant under $U(1)_Y$. The consequences of these results are that the $n=2$ and 3 point correlation functions of the $N=4$
gauge-invariant operators which correspond to KK multiplets in AdS
supergravity are given exactly by their tree level expressions, the 4 point correlation functions of such operators are invariant under $U(1)_Y$ and correlation functions with $n\geq 4$ points have non-trivial dependence on the Yang-Mills coupling constant.

{\vfill\leftline{}\vfill
\vskip	10pt
\footnoterule
{\footnotesize
\hoch{\dagger} Research supported in part by
NSF Grant PHY-9411543 \vskip -12pt}
\vskip	10pt
{\footnotesize
\hoch{\ddagger} Permanent Address: \kings \vskip -12pt}}

\pagebreak
\setcounter{page}{1}

In a series of papers it has been suggested that it might be profitable to study Green's functions of certain classes of gauge-invariant operators in four-dimensional superconformal field theories using superspaces which are specially adapted to the superconformal geometry of these classes  \cite{hw1,hw2}. The operators of interest are initially given  as constrained superfields on Minkowski superspace, but by working in appropriately defined superspaces these constraints can be solved explicitly in a rather simple and geometrically natural fashion \cite{hh}. The basic idea is to study the Green's functions of such operators using the superconformal Ward identities considered as differential constraints on multiple products of these superspaces. For the case of $N=4$ SYM there is a natural class of operators
represented by analytic superfields. This class of operators, for the
gauge group
$SU(N_c)$, coincides with the operators which couple to the
Kaluza-Klein multiplets of IIB supergravity on an $AdS_5\times S^5$
background \cite{af} and is therefore of paramount interest in the context of
the Maldacena conjecture relating IIB supergravity on this background
to $N=4$ superconformal field theory \cite{m}.

A step towards finding the full consequences of the superconformal
Ward identities was taken in reference \cite{hw2} where
superconformal invariants were studied. These invariants generalise the usual cross-ratios which arise as ordinary conformal invariants. However, in a recent paper \cite{ken1}, in which  the possible implications of the
additional
$U(1)_Y$ symmetry of IIB supergravity for $N=4$ SYM were discussed, Intriligator observed that the non-nilpotent invariants constructed in reference
\cite{hw2} are  invariant under this $U(1)_Y$ symmetry.  He  discovered that if one assumes that the correlation functions themselves are invariant under this symmetry then one is led to  the extremely strong conclusion
that they are all independent of the SYM coupling $g_{YM}$. Intriligator's argument may be paraphrased as follows: part of the $N=4$ Yang-Mills action is
not invariant under the  $U(1)_Y$ symmetry and hence
differentiating the
action with respect to the coupling constant results in this non-invariant
term among others. As a result, differentiating any correlator,
assumed to be  $U(1)_Y$ invariant, with respect to the coupling leads to a correlator
with the non-invariant term inserted as an additional operator.
However, this new correlator is not  $U(1)_Y$ invariant and so by
assumption vanishes. It then follows that the original correlator
does not depend on the coupling constant.
Intrilligator then  concluded that either a) the strong conclusion
could be right, b) the harmonic superspace formalism is
flawed, or c) there are more invariants than those listed in \cite{hw2}.

Possibility (a) can be ruled out since although
this might be feasible for $n\leq3$ points it is certainly not
 true for
$n\geq4$ points as explicit calculations have shown \cite{eetal,grps}. We shall comment briefly below on option (b), but the main point of this
note is to argue that possibility (c) holds. This is bound to
be the case since the action contains a $U(1)_Y$ non-invariant piece
which  will lead to non-invariant vertices in the Feynman rules and so to non-invariant Green's functions.
In fact, we show that there is almost certainly a 5-point invariant
whose leading term behaves like $\lambda^4$, where the odd coordinates
of analytic superspace are denoted $\l,\p$. The invariants given in \cite{hw2}
all involve these variables in the combination $\l\p$. In $N=2$ such
a dependence of the invariants on the odd coordinates is dictated by $R$-symmetry, but in $N=4$
it is not because the $R$-symmetry group is $SU(4)$ and not $U(4)$.
Nevertheless, the $\bbZ_4$ centre of $SU(4)$ does place restrictions on the way that the odd variables appear in invariants. If $e$ denotes the generator of ${\bbZ}_4$, where $e^4=1$, one finds that $\l\rightarrow e\l$ whereas $\p\rightarrow\bar e\p$. As a result the odd
variables can only appear in  combinations of the form $\l^p\p^q$
where $q=p$ mod
$4$ \cite{hw1}.
\par
The existence of a $\l^4$ invariant for 5 points explicitly
invalidates Intriligator's argument concerning the non-dependence of
the correlation functions on $g_{YM}$ for $n\geq4$ points. However, we
shall also show that there are no nilpotent invariants for $n\leq4$
points so that all possible 4 point invariants are included correctly
in \cite{hw1}. Thus we can apply Intriligator's
construction to prove the conjectured non-renormalisation theorem for
$n=3$ (and 2) points. The general form of the 3 point functions of analytic operators was given in \cite{hw1} and discussed in more detail in \cite{hsw}; it was shown in \cite{min} that the AdS
supergravity amplitudes and the appropriately normalised free SYM
correlation functions agree and it was conjectured that this might be
true for all values of the 't Hooft coupling $g_{YM}^2N_c$ and even
for all $N_c$. Subsequently this was confirmed to first non-trivial
order in perturbation theory \cite{dzf}. Moreover, there is an anomaly argument
which establishes the strong form of this conjecture for the
correlator of three supercurrent multiplets \cite{hsw} and an alternative argument based on the Adler-Bardeen theorem \cite{gk, dzf2}. One of the main results of this paper
is that the strong conjecture holds for all 3 point functions of
analytic operators.
\par
Of course it is still possible the option (b) holds, namely that
the analytic harmonic superspace  formalism is in some way
flawed. One  concern is that the
underlying SYM multiplet is on-shell, and indeed it is true that the
analyticity of the composite operators under consideration depends on
the field equations of the underlying Yang-Mills fields being
satisfied. A consequence of the on-shell nature of the formalism is
that it is almost impossible to check analyticity directly in $N=4$
perturbation theory. Nevertheless, if one considers the $N=4$ theory
as an $N=2$ theory consisting of a vector multiplet and a
hypermultiplet, both transforming under the adjoint representation of
the gauge group, it is possible to carry out perturbative calculations
in an off-shell
$N=2$ harmonic superspace formalism and it has been verified that
analyticity does indeed hold
for correlation functions of hypermultiplet composites in low orders
in perturbation theory \cite{hsw,eetal}. The present situation is therefore that
analyticity in the $N=4$ formalism should be regarded as an
assumption, but that it is supported by the checks in $N=2$
perturbation theory that have been carried out so far.

We briefly recall the analytic superspace formalism.
$N=4$ analytic superspace $\bbM$ has coordinates
\be
X=\left( \ba{ll} x^{\a\adt} & \l^{\a a'} \\
\p^{a\adt} & y^{a a'} \ea\right)
\ee
where each index can take on 2 values. The even coordinates  $x$ and
$y$ are coordinates for complex spacetime and the internal space
$S(U(2)\xz U(2))\bsh SU(4)$ respectively. The odd coordinates $\l$ and
$\p$ number 8 in all, half the number of odd coordinates of $N=4$
super Minkowski space. An infinitesimal superconformal transformation
takes the form
\be
\d X= \cV X=B + AX + XD + XCX
\ee
where each of the parameter matrices is a $(2|2)\xz (2|2)$
supermatrix and where
\be
\d g=\left( \ba{ll} -A & B \\ -C & D\ea\right) \in \gs\gl(4|4)
\ee
One can show that the central elements in the
superalgebra $\gs\gl(4|4)$ do not  act on $\bbM$ so that one really has
an action of the superalgebra
$\gp\gs\gl(4|4)$.
\par

The gauge-invariant operators are $A_q=\tr (W^q)$ where $W$ is  the
$N=4$ SYM field strength tensor which takes its values in the Lie
algebra $\gs\gu(N_c)$ of the gauge group. These operators transform as
\be
\d A_q=\cV A_q + q\D A_q
\ee
where $\D=\str (A+XC)$. A correlation function of such operators
\be
G(X_1,\ldots X_n)=<A_{q_1}(X_1)\ldots A_{q_n}(X_n)>
\ee
should satisfy the Ward identity
\be
\sum_{i=1}^n (\cV_i + q_i\D_i) G=0
\ee
Such a correlation function, if it does not vanish at zeroth  order in
the odd variables, can be written in the form
\be
G={\rm prefactor} \xz F,
\ee
where the prefactor is a function of the ``propagators''
\be
g_{ij}={\sd} X^{-1}_{ij}={\hat y^2_{ij}\over x^2_{ij}},
\ee
with $X_{ij}=X_i-X_j$ and $\hat y_{ij}=y_{ij}-\p_{ij}x^{-1}_{ij}\l_{ij}$,
which absorbs the charges of the operators and which is analytic in the internal bosonic coordinates. The function $F$ is therefore a function of superconformal invariants.

In \cite{hw2} a large number of superinvariants was found, but,  as pointed
out in \cite{ken1}, they all depend on the odd variables in the combination
$\l\p$. They are thus invariant under $PGL(4|4)$ and not just
$PSL(4|4)$. To examine whether this list is complete or  not we shall look for
nilpotent invariants using the
supersymmetry Ward identities in a straightforward manner. Suppose $F$ is a nilpotent invariant,
then $F=F_o + $ higher order in $\l,\p$, where $F_o$ is itself
nilpotent and has a fixed power of the odd variables. To rule out the
existence of any such $F$ for a given number of points it is therefore sufficient to show that all possible leading terms $F_o$
vanish. The superconformal Ward identities must hold order by
order in powers of the odd variables $\lambda$ and $\p$ so that the action of the Ward
identity on $F_0$ at  lowest order is obtained when 
the  Ward identity operator is linearised appropriately with respect to $\l$ and $\p$.
These truncated  superconformal Ward  identities
involve the following simplified (linearised) superconformal Killing vectors:
\bea 
\cV_{\a\adt}&= &{\del\over\del x^{\a \adt}} \\
\cV(D)&=& x^{\a\adt}\del_{\a\adt} + {1\over2}\l^{\a a'}\del_{\a a'}+{1\over2}\p^{a\adt}\del_{a\adt} \\
\cV^{\a\adt}&=& x^{\b\adt}x^{\a\bdt}\del_{\b\bdt} + x^{\b\adt}\l^{\a b'}\del_{\b b'} + \p^{b \adt} x^{\a \bdt}\del_{b \bdt}
\eea
corresponding to translations, dilations and conformal boosts
as well as three similar equations with $x$ and $y$ interchanged.
The linearised Killing vectors for ordinary $(Q)$ supersymmetries and  special ($S$)
supersymmetries are given
respectively by
\begin{eqnarray}
\cV_{\a a^\prime}&=&{\del\over\del \l^{\a a'}}\\
\la{supl}
\cV_{\a}^{a}&=&y^{a a'}{\del\over\del \l^{\a a'}}\\
\la{supn}
\cV_{a'}^{\adt}&=&x^{\a\adt}{\del\over\del \l^{\a a'}}\\
\la{ssupl}
\cV^{a\adt}&=& x^{\a\adt} y^{a a'}{\del\over\del \l^{\a a'}}
\la{ssupn}
\end{eqnarray}
together with a similar set with $\l$ replaced by $\p$.
The above superconformal Killing vectors can be read off from
the full superconformal Killing vectors given in \cite{hw1}. The Ward identity also involves
the quantity $\D$; however, except for dilations and conformal boosts in $x$ and $y$ this
term vanishes when linearised. 

Translational symmetries of the form of \eq{t} hold for all of the coordinates, so that we can conclude immediately that an invariant can only depend on the differences $X_{ij}:=X_i-X_j$.
\par
In order to solve  equation \eq{ssupl} we change variables from the
differences $\l_{ij}$ to the variables
\be
\l_{123}:= x_{12}^{-1}\l_{12}-x_{23}^{-1}\l_{23},\ \
\l_{234}:= x_{23}^{-1}\l_{23}-x_{34}^{-1}\l_{34}, \ldots
\la{l1}\ee
 and
\be
\l_c= \sum _i x_{i i+1}^{-1}\l_{i i+1}
\ee
It is straightforward to verify that the variables of the
first equation are inert under the transformation induced
in equation \eq{ssupl} and so this equation implies that
the $F_0$ part of the Green's functions does not depend on $\l_c$.
\par
Equation \eq{supn} can be solved in a similar manner. We introduce the
variables
\be
\l_{1234}:=(x_{12}^{-1}
y_{12}-x_{23}^{-1}y_{23})^{-1}\l_{123}-(x_{23}^{-1}
y_{23}-x_{34}^{-1}y_{34})^{-1}\l_{234}, \ldots
\la{l2}\ee
which are inert under the transformation induced by \eq{supn} and the
final variable which is the sum of terms over all points of terms
of this form. Equation \eq{supn}  then implies that $F_0$ only depends
on
$\l_{1243}\ldots$.
\par
Finally,  we turn to equation \eq{ssupn} which is the most complicated
supersymmetry transformation. It implies that $F_0$ depends on only
$\l_{12345}$  where $\l_{12345}$ is given by
\be
\l_{12345}=A_{1234}^{-1}\l_{1234}- A_{2345}^{-1}\l_{2345}, \ldots
\la{l3}\ee
where $A_{1234}$ is obtained by substituting in  the
supersymmetry Ward identity of equation \eq{ssupn} to bring it to the
form
\be
\left(A_{1234}{\del\over
\del\l_{1234}}+ A_{2345}{\del\over\del\l_{2345}}\right)F_o=0
\ee
Explicitly
\bea
A_{1234}&=&\Big( y_{12}z_{123}^{-1} x_{12}^{-1}(x_1 + x_2)-y_{23} z_{123}^{-1} x_{23}^{-1}(x_2 + x_3)\nonumber\\
&\phantom{=}& + (y_1 +y_2) z_{123}^{-1} x_{12}^{-1} x_{12} -(y_2 + y_3)z_{123}^{-1} x_{23}^{-1} x_{23}\Big) -\Big({\rm same\ with\ } (123)\rightarrow (234)\Big)
\eea
where
\be
z_{123}:=y_{12} x_{12}^{-1}-y_{23} x_{23}^{-1}
\ee
Moreover it is possible to invert
$A_{1234}$ and thus obtain a rational expression for $\l_{12345}$. This
expression is rather complicated, however, so that we shall not give
it here.
We find essentially identical results with $\l$ replaced by $\p$
if we use the corresponding Ward identities.
\par
The conformal boost Ward identity and a  similar
identity  in the internal space are now automatically satisfied as
a result of taking the anti-commutators of the $Q$ and $S$ supersymmetry transformations
leaving only the Ward Identities for spacetime and internal dilations. However, these are easily solved and
just determine the overall power of $x$ and $y$ respectively.
\par
Let us carry out a count of the spinor variables that a $n$ point invariant or
Green's functions can depend on. Initially,
an  invariant depends on all the spinor coordinates $\l_i,\
\p_i,\   i=1,\ldots n$ that is $4n$  $\l$'s and $4n$ $\p$'s. The
translational supersymmetries \eq{supl} imply
that it can only depend on differences,  that is, on the $4(n-1)$
$\l_{i i+1}$'s
 with a similar result for $\p$'s. In a similar way, equations
\eq{supn},
\eq{ssupl} and \eq{ssupn} imply that an invaraint can only
really depend on $4(n-4)$ spinors of equation
\eq{l3} with a similar result for $\pi$'s.
Hence for Green's functions with $n\le 4$ points there are in effect no
available spinors with which to form nilpotent invariants. 

We now
discuss in more detail these consequences of the superconformal Ward identities  for Green's functions with a small number of points.
 The simplest example is at 2 points.
$F_o$ can only  depend on
$X_{12}$.  Then \eq{sup} implies in this case that
\be
x^{\a\adt}_{12}{\del\over\del \l_{12}^{\a a'}}F_o=0
\ee
and this in turn implies that $F_o$ cannot depend on $\l_{12}$,  or,
by a similar argument $\p_{12}$. Hence there can be no nilpotent 2
point invariants.
\par
Three-point correlation functions have  been studied in \cite{hsw}. If the sum
of the charges, $Q=\sum_i q_i$, is even, the result is
\be
<A_{q_1} A_{q_2} A_{q_3}>=C_{q_1 q_2 q_3}(g_{12})^{k_1} (g_{23})^{k_2}
(g_{31})^{k_3}
\ee
where $C$ is a constant and where
\bea
k_1={1\over2}(q_1 + q_2 - q_3) \\
k_2={1\over2}(q_2 + q_3 - q_1) \\
k_3={1\over2}(q_3 + q_1 - q_2)
\eea
This solution is unique up to the constant involved because there
are no 3 point invariants. There are clearly no non-nilpotent 3 point
invariants because the leading term of such an invariant would be
either a 3 point spacetime invariant or a 3 point internal space
invariant and there are no such objects. To examine the existence of
nilpotent invariants we use the above method. This time
\eq{ssupl} implies that $F_o$ can depend on
$\l_{123}$
and a similar $\p$ variable, but  \eq{supn} shows that this dependence
must be trivial. Thus there are no 3 point invariants.  An identical
argument can be used to show that the 3 point functions with odd total
charge, and which consequently vanish to leading order, must in fact
vanish to all orders. Without loss of generality we can take such a 3
point function to be specified by charges $q_1,q_2,q_3+1$, where
$\sum_i q_i$ is again even. It can be written
\be
<A_{q_1} A_{q_2} A_{q_3+1}>=(g_{12})^{k_1} (g_{23})^{k_2} (g_{31})^{k_3}F
\ee
where $F$ is nilpotent and satisfies
\be
(\sum_i (\cV_i) +\D_3)F=0
\ee
However, since the $\D_3$ term vanishes  for linear $S$ and $Q$
supersymmetries we can immediately deduce that $F_0$ and therefore also $F$ vanish.

For 4 points, equations \eq{supl}  and
\eq{ssupl} imply that  the leading term, $F_o$, of a putative
nilpotent invariant $F$ depends only on the odd variables
of equation \eq{l1}
and a similar pair of $\p$ variables, while \eq{supn} implies that $F_o$ should only depend on
the $\l_{1234}$ spinor of equation \eq{l2}
as well as a similarly defined $\p_{1234}$. Implementing  finally the
non-linear $S$ supersymmetry \eq{ssupn}  we find that this
dependence is actually trivial, and so we conclude that there can be
no nilpotent invariants for 4 points. This means that the 4 point
invariants are determined by their leading, purely bosonic terms.
These are conformal invariants in
$x$ and $y$, that is, cross-ratios of the $x$ and $y$ differences.
There are two independent such variables in both sectors for 4 points
and so 4 independent invariants altogether. They may be expressed in
terms of the superinvariants given in \cite{hw2}; for example, one could take
as a basis set two super cross ratios of the form
\be
{{\sd}\, X_{14}\, {\sd}\, X_{23}\over {\sd}\, X_{12}\, {\sd}\, X_{34}}
\ee
and 2 supertraces of the form
\be
{\str}\, (X_{12}^{-1}X_{23} X_{34}^{-1} X_{41})
\ee
An important conclusion of this analysis is that the invariants listed in \cite{hw2} are indeed complete for 4 points, so that these invariants are in fact invariant under $PGL(4|4)$ and not just $PSL(4|4)$.
\par
Applying a similar argument to the 5 point case we  find that the
leading term, $F_o$, of a 5 point nilpotent invariant can only
 depend
on the odd variables
$\l_{12345}$ of equation \eq{l3} and a similar $\p_{12345}$ variable
At first sight this expression seems to involve  dependence on $x_i$
and not just the differences $x_{ij}$, but in fact this turns out not
to be the case although  $\l_{12345}$
as defined in \eq{l3}  has  complicated
$x$-dependent singularities in the $y_{ij}$ which are not allowed in
correlation functions. However, these can be removed by simply
multiplying through by the denominator, a procedure which does not
affect the supersymmetry analysis given above at this lowest order.
Although the Green's functions are not necessarily invariant under
$U(1)_Y$ they are invariant under $\bbZ_4$ and so the
leading term of a five-point nilpotent invariant can
 be of the form
$\l^4$ or of the form
$\p^4$. From
$\l_{12345}$ of \eq{l3} we can construct the leading term of a nilpotent
5 point invariant of the form
$\l_{12345}^4 f(x,y)$, for some appropriate function $f$ of the
$x_{ij}'s$ and the
$y_{ij}'s$. The dilation Ward identity and a similar
identity for the internal variable $y$ imply that the dependence of $f$ on $x$ and $y$ is schematically of the form $x^{-2}y^{-2}$.  Although we
have no proof at present that this leading term can be extended to a
full 5 point superinvariant it seems highly probable that such an extension
exists. We note, in particularly, that such an invariant is not
$U(1)_Y$ invariant as it does not depend on the odd variables as a
power series in $\l\p$.

To obtain the consequences of the above results on  nilpotent
invariants we briefly review Intriligator's argument concerning the
dependence of correlation functions on the coupling. To do this in our
formalism we first note that the supercurrent $T=\tr (W^2)$ is also
the (on-shell) Lagrangian mulitplet. In (real) Minkowski superspace
the supercurrent multiplet is $T_{ij,kl}:=\tr (W_{ij}W_{kl})_{20}$,
where $i,j=1,\ldots 4$ now denote $SU(4)$ indices and where the
subscript $20$ indicates that the real twenty-dimensional
representation is to be projected out from the product by imposing the condition
$\epsilon ^{ijkl}T_{ij,kl}=0$. Reality means
that
\be
T^{ij,kl}:={1\over4}\e^{ijmn} \e^{klpq} T_{mn,pq}=\bar T^{ij,kl}
\ee
If we set
\be
T'^{ij,kl}:=T^{ik,jl}+T^{jk,il}
\ee
then $T'$ is symmetric on both pairs of indices,  symmetric under the
interchange of the pairs and vanishes on symmetrisation over any 3
indices. Using $T'$ we can form a complex superaction \cite{hst}
\be
{\cS}=\int d^4x\, D_{ij} D_{kl} T'^{ij,kl}
\ee
where
\be
D_{ij}:=D_{\a i} D^{\a}_j=D_{ji}
\ee
The invariance of $\cS$ under supersymmetry follows from  the
constraints 
\be
D_{\a i}T'^{jk,lm}=2\d_i^{(j}\L_{\a}^{k),lm}=2\d_i^{(l}\L_{\a}^{m),jk}
\ee
where
\be
\L_{\a}^{j,kl}:={1\over4}D_{\a i}T'^{ij,kl}
\ee
and
\be
\bar D_{\adt}^{(i}T'^{jk),kl}=0
\ee
These constraints are themselves a consequence of the constraints satisfied be the underlying field strength $W_{ij}$:
\be
\nab_{\a i} W_{jk}=\nab_{\a [i} W_{jk]}
\ee
where $\nab_{\a i}$ is the gauge-covariant spinorial derivative.
It is straightforward to compute that
\be
{\cS}=\int d^4x\ (-{1\over4} F_{\a\b} F^{\a\b} + \ldots)
\ee
$F_{\a\b}$ being the self-dual part of the  Minkowski space
Yang-Mills field strength tensor.

The analytic supercurrent $T$ is related  to $T_{ij,kl}$ by means of
the
$SU(4)$ harmonic variables $(u_r{}^i,\ u_{r'}{}^i) \in SU(4),\
r=1,2,\,r'=3,4$.
\be
T={1\over4}\e^{rs}\e^{tu} u_r{}^i u_s{}^j u_t{}^k u_u{}^l T_{ij,kl}
\ee
We can therefore rewrite the superaction as a harmonic superaction
\be
{\cS}=\int d\m\ T
\ee
where
\be
d\m := d^4x\,du\, (D')^4
\ee
with
\be D'\sim D_{\a r'}:=u_{r'}{}^i D_{\a i}
\ee
and where $du$ denotes the standard  invariant measure on the coset
$S(U(2)\times U(2))\bsh SU(4)$. Using this formalism we may then write
the on-shell action  as
\be
S={\rm Im}\left( \t\int d\m\,T\right)
\ee
where $\t$ is the coupling
\be
\t:={\th\over 2\p} +{4\p i\over g_{YM}^2}
\ee
Note that, in terms of the odd variables $\l,\p$  the measure $d\m$
essentially contains the factor $d^4\l$.

If we differentiate a correlation function  with respect to the
coupling we get
\be
{\del\over\del \t}<{\cA}_{q_1}\ldots {\cA}_{q_n}>\sim\int d\m\,<{\cT}{\cA}_{q_1}\ldots {\cA}_{q_n}>
\ee
For example, for a 3 point correlator we have
\be
{\del\over\del \t}<{\cA}_{q_2}{\cA}_{q_3}{\cA}_{q_4}>\sim\int
d\m_1\,<{\cT}(1){\cA}_{q_2}{\cA}_{q_3}{\cA}_{q_4}>
\ee
where ${\cA}_q=(g_{YM})^{-q}A_q$. The differentiation of the path integral expression for the correlation function sees these explicit factors of $g_{YM}$, but these terms are cancelled by a term arising from the differentiation of the action which counts the powers of fields in the operators. The residual term from differentiating the action is then the on-shell integrated action. Now the integral on the right projects out the $(\l_1)^4$ term in the 4
point correlator. Since this correlator depends on the odd variables only
through $\l\p$ the result would have to have the form $\p^4\times$ a power
series in $\l\p$. However, the 3 point correlator also depends only on
$\l\p$ and therefore has no such terms. Thus the integral is zero and we
conclude that all 3 point correlators have trivial dependence on the
coupling.

Such a conclusion will not hold for 4 point correlators, however, provided
that the putative 5 point nilpotent invariant discussed above exists. The
leading term of this invariant can be expressed in terms of the differences
$\l_{12},\l_{13},\l_{14},\l_{15}$ and so gives a contribution of the
integrand of the form $(\l_1)^4$ which moreover has no $\p's$. Therefore
one cannot conclude that 4 point functions, or indeed $n$ point functions
with $n\geq 4$, should have trivial coupling constant dependence.

To conclude, we have shown that the conjectured non-renormalisation theorem
for 2 and 3 point correlation functions in $N=4$ Yang-Mills theory holds exactly.
This lends further support to the Maldacena conjecture to add to the results obtained using instanton techniques \cite{inst}. We have also shown that the 4 point invariants are correctly listed in \cite{hw2}, so that the 4 point correlation functions of analytic operators are actually invariant under $U(1)_Y$. We have also seen that it seems likely that there is a 5 point
invariant whose leading term behaves like $\l^4$. Such an invariant would
not be invariant under $U(1)_Y$ and its existence would imply that $n$
point functions for $n\geq 4$ do not depend trivially on the coupling.

{\bf Note added} In a recently posted paper \cite{ken2} an argument for the 3-point non-renormalisation theorem was given based on a conjecture concerning the behaviour of the OPE. The authors of this paper have suggested, on the basis that it leads to $U(1)_Y$ invariant correlation functions, that the analytic superspace method is flawed. However, as we have shown above, there are now good grounds to suppose that this will not be true for $n\geq 5$ points. These authors also suggest that the analytic superspace formalism is not capable of accommodating the long supermultiplets of the theory. We believe that this is not the case; this point will be discussed further elsewhere.

\end{document}